\documentclass[preprint,3p,times,twocolumn]{elsarticle}
\usepackage{amsmath,amssymb, amsfonts}
\usepackage{mathptmx}
\usepackage{graphicx}
\usepackage{color}
\usepackage{physics}
\usepackage{booktabs}
\usepackage{array}
\usepackage{listings}
\usepackage{soul}
\usepackage{xcolor}
\usepackage[bookmarksopen, bookmarksdepth=2,breaklinks=true]{hyperref}
\hypersetup{
  colorlinks,
  linkcolor={red!50!black},
  citecolor={blue!50!black},
  urlcolor={blue!80!black}
}

\newcommand{\mat}[1]{\mathbf{#1}}
\newcommand{\gr}{\mat{G}}
\newcommand{\vk}{\vec{k}}
\newcommand{\vR}{\vec{R}}

\newcommand{\mH}{\mat{H}}
\newcommand{\mG}{\mat{G}}

\newcommand{\matp}{\mat{p}}

\let\oldhref\href
\renewcommand{\href}[2]{\oldhref{#1}{\hbox{#2}}}
\begin{document}

\title{TB2J: a python package for computing magnetic interaction parameters}
\author[1,2,3]{Xu He\corref{cor1}}
\ead{mailhexu@gmail.com}

\author[3]{Nicole Helbig}

\author[2,3]{Matthieu J. Verstraete}

\author[1]{Eric Bousquet}

\cortext[cor1]{Corresponding author}
\address[1]{Physique Th\'eorique des Mat\'eriaux, Q-MAT, CESAM, Universit\'e de
  Li\'ege, B-4000 Sart-Tilman, Belgium}
\address[2]{Catalan Institute of Nanoscience and Nanotechnology (ICN2), CSIC, BIST, Campus UAB, Bellaterra, Barcelona, 08193, Spain}
\address[3]{Nanomat, Q-Mat, CESAM, and European Theoretical Spectroscopy Facility, Universit\'e de Li\`ege, B-4000 Li\`ege, Belgium}

\begin{abstract}
We present TB2J, a Python package for the automatic computation of magnetic interactions, including exchange and Dzyaloshinskii-Moriya, between atoms of magnetic crystals from the results of density functional calculations. The program is based on the Green's function method with the local rigid spin rotation treated as a perturbation. As input, the package uses the output of either Wannier90,  which is interfaced with many density functional theory packages, or of codes based on localised orbitals. A minimal user input is needed, which allows for easy  integration into high-throughput workflows. 
\end{abstract}

\maketitle
\section{Introduction}
First-principles simulations of magnetic materials have attracted strong interest in the last decade due to an increase in precision provided by density functional theory (DFT) codes for strongly correlated magnetic atoms and also because of the developments in  spintronics applications~\cite{hirohata2020, baltz2018, bhatti2017, joshi2016, lu2016,zutic2004}. Understanding the complex microscopic origin of the magnetic interactions often requires a reduction of the full many-body electronic interactions to effective Hamiltonians, the two most important ones being the Heisenberg and the Hubbard Hamiltonians~\cite{fazekas1999}. The parameters of these Hamiltonians are fit to DFT data to give access to an easier understanding of the magnetic interactions and to allow for simulations of larger systems and/or with dynamics.
Novel sensing, storing and computing technologies have been proposed using spin waves~\cite{Chumak2015} and skyrmions~\cite{Kiselev2011,Fert2013}. Theoretical models have consistently opened new vistas and explained experimental findings in complex new chemistries, geometries, and heterostructures. Reliable ab initio calculations of the effective Hamiltonian parameters is crucial for predictive and quantitative simulations.
Fitting these parameters is, however, often very cumbersome and necessitates a case-by-case construction~\cite{Xiang2013}.

For the Heisenberg Hamiltonian, one of the most common fitting procedures is through total energy calculations or energy mapping analysis~\cite{Xiang2013}. This approach necessitates the calculation of the total energies of different magnetic configurations. The parameters of the Heisenberg Hamiltonian are then fit to these energies under the supposition that the change in energy is only related to the magnetic interactions. This requires to have at least as many calculated magnetic configurations as the number of parameters of the Hamiltonian (though more magnetic configurations are usually necessary to converge the fit), which often requires the use of large supercells. This simple method gives good results in some cases but the Heisenberg model itself can break down if the chosen magnetic configurations deviate too much from the ground-state one. This can happen, e.g., if a magnetic phase, often the ferromagnetic (FM) one, closes the band gap of an insulating antiferromagnetic (AFM) ground-state. Also, the method becomes unsuitable if the supercell needed to probe all of the pertinent magnetic configurations is too large to be handled by DFT. For example, if certain configurations show a delocalized picture for the electrons they must be excluded from the fit, which typically leads to an increase in the size of the supercell needed to provide a sufficiently large number of configurations.

Another method to determine the parameters is the Generalized Bloch Theorem (GBT) \cite{herring1966magnetism, sandratskii1986} 
which adapts the boundary conditions on the spin orientation to account for a spin spiral. From the change in total energy one can extract the magnetic interaction parameters. As yet another alternative, Density Functional Perturbation Theory (DFPT) has been extended to magnetic field perturbations by Savrasov \cite{Savrasov1998_prl_81_2570} and yields the susceptibility and the magnetic exchange as a sub-product. This perturbation has been implemented in abinit \cite{Romero2020_jcp_152_124102} and quantum Espresso \cite{Cao2018_prb_97_024420}, and has also been combined with atomic displacements within non-collinear formalisms \cite{Ricci2019_prb_99_184404}. For several magnetic ions in the unit cell, a local perturbation DFPT scheme is presented in Ref.~\cite{Phillips2013}.

A different procedure, which avoids the problems of the total energy mapping, employs Green's functions by taking the local spin rotation as a perturbation, as proposed in the seminal work of Liechtenstein, Katsnelson, Antropov and Gubanov (LKAG) \cite{Liechtenstein1987}. The Green's function method allows to determine the Heisenberg magnetic interaction parameters from the ground-state solution of the system, regardless of whether it is FM or AFM. This method also allows access to a band-by-band decomposition of the different magnetic interactions, and it is easier to automatize than the total energy mapping \cite{Steenbock2018}. The method was also extended to correlated systems in Refs.\ \cite{Katsnelson2000, Katsnelson2010}. By taking into account relativistic effects (spin-orbit coupling), the Dzyaloshinskii-Moriya interaction (DMI)  \cite{Solovyev1996, Antropov1997, Mazurenko2005, Mankovsky2017} and the magnetic anisotropy can be calculated as well \cite{Katsnelson2000, Katsnelson2010}. The method has been extended to orbital-spin and orbital-orbital magnetic interactions \cite{Secchi2015,Secchi2016}. Higher order terms in the Hamiltonian, like the four-spin interaction or the biquadratic term, were also calculated through this method \cite{Lounis2010, Szilva2013}. The LKAG method of computing magnetic interactions has been initially implemented with the Korringa-Kohn-Rostoker Green’s function (KKR-Green) \cite{Yavorsky2006, Ebert2009} and tight-binding linear muffin-tin methods \cite{Anderson1984}, and has proven to be quite efficient. Its extension to other basis sets popular for DFT calculations will strongly broaden its usage.

Constructing the Green's function has proven to be cumbersome for non-localized basis sets (e.g.\ plane waves), but this difficulty is greatly reduced by the use of Wannier functions (WF) as done by Kokorin et al.\ in \cite{Korotin2015}. The WFs can be constructed from first principles with the widely used open source code Wannier90 \cite{Mostofi2008,Pizzi2020}.  Wannier90 has been interfaced with a large number of DFT codes, including Abinit \cite{Gonze2009,Gonze2020}, Quantum Espresso \cite{Giannozzi2017}, Siesta \cite{Soler2002}, VASP \cite{Kresse1996,Kresse1999}, Wien2K \cite{Blaha2001}, Fleur \cite{Blugel2006}, Octopus \cite{Andrade2015, Tancogne2020}, OpenMX \cite{Ozaki2005}, GPAW \cite{Enkovaara2010}, ELK \cite{elk}, and many others. 
Several codes, including exchange.x \cite{Korotin2015}, nojij \cite{Oroszlany2019}, and Jx \cite{Yoon2020}, exist which calculate magnetic exchange parameters from Wannier functions or linear combination of atomic orbitals (LCAO) DFT results. However, to the best of our knowledge, DMI and anisotropic exchange from non-relativistic effects have not been integrated yet.

Here, we present a Python package, TB2J, that allows for automatic and systematic calculations of the parameters of a Heisenberg Hamiltonian through the Korotin approach \cite{Korotin2015}. The script can compute the isotropic exchanges, the anisotropic exchanges, and the DMI from the output of the Wannier90 code or from a LCAO Hamiltonian (Siesta and OpenMX through the SISL package \cite{zerothi_sisl}, and GPAW) following the scheme proposed in Refs.\ \cite{Liechtenstein1987, Antropov1997, Szilva2013}. TB2J is designed with the goal to minimize the number of inputs and actions from the user. In most cases, only the paths of the Wannier or DFT related files and the species of magnetic atoms are mandatory. Several types of output files are generated which can then be used in spin dynamics and Monte Carlo codes. TB2J's API is implemented in an abstract manner that eases interfacing with new tight-binding-like Hamiltonians or spin dynamics simulation.

We first present the general methods used in TB2J and then describe how these methods are implemented. Next, we exemplify the usage of TB2J by applying it to, BCC Fe, HCP Co bulk, SrMnO$_3$, BiFeO$_3$, and La$_2$CuO$_4$ crystals. In the end we discuss the advantages and limitations of this method. 

Throughout this paper, vectors are denoted with the $\vec{\ }$ notation while matrices are represented with bold characters.

\section{Formalism and algorithms}

The main idea of the method is to perturb a localized spin, in both the Heisenberg spin model and the DFT electronic model. For the latter, the rigid spin-rotation perturbation is done within the single-particle Green's function method. The Heisenberg parameters are then mapped to the electron expressions~\cite{Liechtenstein1987}. 
We will consider the quantization axis along $z$ in the following.
\subsection{Heisenberg model}

The Heisenberg Hamiltonian contains four different parts and reads as 
\begin{eqnarray}
 \nonumber
\lefteqn{E =  -\sum_i K_i (\vec{S}_i\cdot \vec{e}_i)^2} \\
\nonumber
&&-\sum_{i \neq j} \biggl[ J^{iso}_{ij} \vec{S}_i\cdot\vec{S}_j \hspace{0.8cm} \\
 \nonumber
&&+ \vec{S}_i \mat{J}^{ani}_{ij} \vec{S}_j \\
&&+ \vec{D}_{ij} \cdot \left( \vec{S}_i\times\vec{S}_j\right) \biggl],  
 \label{eq:bilinear}
\end{eqnarray}  
where the first term represents the single-ion anisotropy (SIA), the second is the isotropic exchange, and the third term is the symmetric anisotropic exchange, where $\mat{J}^{ani}$ is a $3\times 3$ symmetric tensor. The final term is the DMI, which is antisymmetric. Importantly, the SIA is not accessible from Wannier90 as it requires separately the spin-orbit coupling part of the Hamiltonian \cite{Solovyev1995}. However, it is readily accessible from constrained DFT calculations \cite{weingart2012}. 
We note that there are several conventions for the Heisenberg Hamiltonian, here we take a commonly used one in atomic spin dynamics: we use a minus sign in the exchange terms, i.e.\ positive exchange $J$ values favor ferromagnetic alignment. Every pair $ij$ is taken into account twice, $J_{ij}$ and $J_{ji}$ are both in the Hamiltonian. The spin vectors $\vec{S}_i$ are normalized to 1, so that the parameters are in units of energy. The other commonly used conventions differ in a prefactor ${1/2}$ or a summation over different $ij$ pairs only. The conversion factors to other conventions are given in Table \ref{tab:covention}. For other conventions in which the spins are not normalized, the parameters need to be divided by $\abs{S_i} \abs{S_j}$ in addition.

From the total energy due to the spin interactions, Eq.\ (\ref{eq:bilinear}), we obtain the following variation with respect to the  $\vec{S_i}$ and $\vec{S_j}$  

\begin{equation}
\begin{aligned}
\delta E_{ij} =&  -2 J^{iso}_{ij}\delta \vec{S_i}\cdot \delta\vec{S}_j \\
& - 2 \delta \vec{S}_i \mat{J}_{ij}^{ani}\delta \vec{S}_j\\
& - 2 \vec{D}_{ij}\cdot (\delta \vec{S_i}\cross \delta\vec{S}_j)
\end{aligned}
\label{eq:dEdS2}
\end{equation}


\begin{table}
    \centering
    \begin{tabular}{ccc}
    \toprule
    &  Convention & $\Tilde{J}_{ij}$ \\
    \midrule
         & $-\frac{1}{2} \sum\limits_{ij} \Tilde{J}_{ij} \vec{S}_i \cdot \vec{S}_j$ & $2J_{ij}$ \\ 
         & $- \sum\limits_{<ij>} \Tilde{J}_{ij} \vec{S}_i \cdot \vec{S}_j$ & $2J_{ij}$  \\
        & $ \frac{1}{2}\sum\limits_{ij} \Tilde{J}_{ij} \vec{S}_i \cdot \vec{S}_j$ & $-2J_{ij}$  \\
        & $\sum\limits_{<ij>} \Tilde{J}_{ij} \vec{S}_i \cdot \vec{S}_j$ & $-2J_{ij}$  \\
             & $\sum\limits_{ij} \Tilde{J}_{ij} \vec{S}_i \cdot \vec{S}_j$ & $-J_{ij}$  \\
    \bottomrule
    \end{tabular}
    \caption{The conversion of $J_{ij}$ to other conventions, where $\Tilde{J}$ is the exchange parameter in that convention. The notation $<ij>$ means a pair of $ij$ without counting it twice. The DMI parameters $\vec{D}$ can be converted in the same way.}
    \label{tab:covention}
\end{table}

\subsection{Tight-binding Hamiltonian and Green's function}

We start from a generalized tight-binding Hamiltonian. The localized basis functions are denoted as $\psi_{im\sigma}(\vec{r})$ with $i$, $m$, and $\sigma$ being the site, orbital, and spin indices, respectively. Due to translation symmetry, the tight-binding Hamiltonian, $\mat{H}$, and the overlap, $\mat{S}$, matrices can be parameterized as
\begin{eqnarray}
  \label{eq:WFTB}
  \mat{H}_{im jm^{\prime}\sigma\sigma^{\prime}}(\vec{R})  &=&
  \bra*{\psi_{im\sigma}(\vec{r})} H \ket*{
  \psi_{jm^{\prime}\sigma^{\prime}}(\vec{r}+\vec{R})}, \\
  \mat{S}_{im jm^{\prime}\sigma\sigma^{\prime}}(\vec{R})  &=& \bra*{\psi_{im\sigma}(\vec{r})} \ket*{ \psi_{jm^{\prime}\sigma^{\prime}}(\vec{r}+\vec{R})},
\end{eqnarray}
where $\vec{r}$ is the position inside the unit cell, $\vR$ is the lattice vector, and $H$ denotes the total Hamiltonian. The overlap matrix $\mat{S}$ reduces to the identity matrix when the basis functions are orthonormal. TB2J can use both non-orthogonal LCAO basis sets and orthogonal Wannier basis sets. Below we discuss only  the results for an orthogonal basis set. The non-orthogonal basis set is discussed in Ref.\ \cite{Oroszlany2019}, where it is shown that the expressions for the exchange parameters are the same as for an orthogonal basis. 

In the following, we drop all orbital and spin indices for simplicity, Hence $\mat{H}_{ij}$ is a sub-matrix of $\mat{H}$ containing all the spin and orbital components for atoms $i$ and $j$. 
We note that in DFT it is possible to do non-collinear calculations with or without SOC (the exchange correlation potential can induce a non-collinear ground state).
If one were to perform a calculation with SOC and the spin constrained to one direction, it would be categorized as collinear. 
In the collinear case, $\mat{H}$ is diagonal in the spin subspace.

The Green's function in reciprocal space is defined as
\begin{equation}
  \label{eq:Greenk}
  \gr(\vec{k}, \epsilon) =  \left(\epsilon \mat{S}(\vec{k}) -\mat{H}(\vec{k})\right)^{-1} \,,
\end{equation}
where $\mat{H}(\vec{k})= \sum_{\vR} \mat{H}(\vR) e^{i \vec{k} \cdot \vec{R}}$, and $\mat{S}(\vec{k})= \sum_{\vR} \mat{S}(\vR) e^{i \vec{k} \cdot \vec{R}}$. The Green's function in real space is obtained using the following expression
\begin{equation}
  \label{eq:GR}
  \gr(\vR,\epsilon) = \int_{BZ}  \gr(\vk, \epsilon)e^{-i \vec{k} \cdot \vec{R}} \ d\vec{k}.
\end{equation}
In the following, we drop the $\vR$ in the atom pair labeled by $i,j, \vec{R}$, e.g.\ $G_{ij}$ means $G_{ij}(\vec{R})$, and $G_{ji}$ means $G_{ji}(-\vec{R})$. From Eq.\ (\ref{eq:P}) all equations are given in real space. 

For each atom $i$, the intra-atomic component of $\mat{H}$ is defined as $\mat{P}_i=\mat{H}_{ii}(\vec{R}=0)$ and is of size $2 N_{orb} \times 2 N_{orb}$. Each $\mat{P}_{i,mm'}$ is a $2\times 2$ matrix in spin, which can be decomposed into its scalar and  vector parts
\begin{eqnarray}
  \label{eq:P}
  \mat{P}_{imm'} &=& p^0_{imm'} \mathrm{I} + \vec{p}_{imm'} \cdot \vec{\mat{\sigma}}, \\
  &=& p^0_{imm'} \mathrm{I} + p_{imm'} \vec{e}_{imm'} \cdot \vec{\mat{\sigma}}, \nonumber
\end{eqnarray}
where $\vec{p}_{imm'}$ is the vector part of $\mat{P}$, (upper case $\mat{P}$ denotes matrices in spin \emph{and} orbitals, lower case $\matp$ is used for matrices in orbitals only) which has the $x$, $y$, and $z$ components $p_{imm'}^x, p_{imm'}^y, p_{imm'}^z$. $\vec{e}_{imm'}$ is the unit orientation vector of $\vec{p}_{imm'}$, and $\vec{\mat{\sigma}}=(\mat{\sigma}_x, \mat{\sigma}_z,\mat{\sigma}_z)$ are the three Pauli matrices. In condensed form, the spin and orbital matrix for site $i$ is now $\mat{P}_{i } = \mat{p}^0_{i}\mathrm{I} + \vec{\matp}_{i} \cdot \vec{\mat{\sigma}}$. We can decompose the Green's function for each inter-site orbital pair $\gr_{im,jm\prime}$ in the same way
\begin{equation}
  \label{eq:T}
    \gr_{im,jm\prime} = G^0_{im,jm\prime} \mathrm{I} + \vec{G}_{im,jm\prime} \cdot \vec{\sigma},
\end{equation}
 where $G^0_{im,jm\prime}$ and $\vec{G}_{im,jm\prime}$ form the $\mat{G}^0_{ij}$ and $\vec{\mat{G}}_{ij}$ matrices, respectively.


\subsection{Magnetic force theorem}

According to the force theorem the total energy variation due to a small perturbation from the ground state coincides with the change of the single-particle energies at fixed ground-state potential
\begin{equation}
  \begin{aligned}
    \delta E &=\int_{-\infty}^{E_F} \epsilon \delta n(\epsilon)\ d\epsilon =-\int_{-\infty}^{E_F}   \delta N(\epsilon)\ d\epsilon  
\end{aligned}
\end{equation}
where $n(\epsilon)=-\frac{1}{\pi}Im \Tr(\mG(\epsilon))$ is the density of states and $N(\epsilon) = -\frac{1}{\pi}Im \Tr(\epsilon-\mH)$ is the integrated density of states. The traces are taken over orbitals only, not spin. Thus, the first order variation of $N$ due to $\delta H$ can be written as
$\delta N(\epsilon)= \frac{1}{\pi}Im \Tr(\delta \mH \mG)$; the second order variation
is $\delta^{2} N(\epsilon)= \frac{1}{\pi}Im \Tr(\delta \mH \mG \delta \mH \mG)$.

Now we use the spin rotation as a perturbation. For the rotation of the spin at site $i$, the change in the energy up to the second order reads as
\begin{equation}
  \label{eq:onerot}
  \delta E^{1spin}_{i}=-\frac{1}{\pi}\int_{-\infty}^{E_F}  Im \Tr( \delta \mH_{i} \mG + \delta \mH_{i} \mG \delta \mH_{i} \mG)\ d\epsilon.
\end{equation}
Similarly, for the rotation of the spins at sites $i$ and $j$, the change in energy is given by
\begin{equation}
  \label{eq:tworot}
  \begin{aligned}
    \delta E^{2spin}_{ij}=-\frac{1}{\pi}\int_{-\infty}^{E_F}  Im \Tr \left[ \right.& \delta \mH_{i} \mG + \delta \mH_{i} \mG \delta \mH_{i} \mG  \\
    + & \delta \mH_{j} \mG + \delta \mH_{j} \mG \delta \mH_{j} \mG \\
    + &  2\delta \mH_{i} \mG \delta \mH_{j} \mG \left. \right ] \ d\epsilon.
\end{aligned}
\end{equation}
The energy variation due to the two-spin interaction is then
\begin{equation}
  \label{eq:twospinint}
  \begin{aligned}
\delta E_{ij}=&\delta E^{2spin}_{ij} - \delta E^{1spin}_i-  \delta E^{1spin}_j  = \\
& -\frac{2}{\pi} \int_{-\infty}^{E_F}  Im \Tr( \delta \mH_{i}
\mG \delta \mH_{j} \mG )\ d\epsilon.
\end{aligned}
\end{equation}
The change of $\mat{H}$ due to the rotation of spin is $\vec{\delta \phi}\cross \vec{\matp}$ with the rotation axis along $\vec{\delta\phi}$ and the angle $\abs{\vec{\delta\phi}}$.

By putting this into Eqn.~\ref{eq:twospinint}, we get
\begin{equation}
\begin{aligned}
\delta E_{ij} &=  -2 [A^{00}_{ij}-\sum_{u=x,y,z}A_{ij}^{uv}]\delta \vec{e_i}\cdot \delta\vec{e}_j \\
& - 2 \sum_{u,v \in x,y,z} \delta e_i^u [A_{ij}^{uv}+A_{ij}^{vu}]\delta e_j^v\\
& - 2 \vec{d}_{ij}\cdot (\delta \vec{e_i}\cross \delta\vec{e}_j)
\end{aligned}
\label{eq:deltaEijelectron}
\end{equation}


in which the $4\times 4$ matrix $\mat{A}_{ij}$ is defined as
\begin{equation}
  \label{eq:defA}
 A_{ij}^{uv} =-\frac{1}{\pi} \int_{-\infty}^{E_F}   \Tr{\mat{p}_i^z\mat{G}_{ij}^u \mat{p}_j^z \mat{G}_{ji}^v}\ d\epsilon,
\end{equation}
where $u, v \in  \{ 0, x, y, z\}$, and the component of $\vec{d}_{ij}$, $d_{ij}^u=\Re(A_{ij}^{0u}-A_{ij}^{u0})$. Comparing Eq.~(\ref{eq:deltaEijelectron}) to Eq.~(\ref{eq:dEdS2}), we can find that the values of the isotropic exchange $J^{iso}$ , the anisotropic exchange $\mat{J}^{ani}$, and the DMI $\vec{D}$ can be expressed as
\begin{align}
J^{iso}_{ij}&=\Im(A_{ij}^{00}-A_{ij}^{xx}-A_{ij}^{yy}-A_{ij}^{zz}) \label{eq:Jiso}, \\
 J_{ij}^{ani,uv} & = \Im(A_{ij}^{uv}+A_{ij}^{vu}) \label{eq:Jani}, \\
  D^{u}_{ij} &=  \Re (A_{ij}^{0u} - A_{ij}^{{u 0}})  \label{eq:DMI},
\end{align}
which is the same with Ref.~\cite{Szilva2013}.


If spin-orbit coupling (SOC) is neglected, $A_{ij}^{0u} = A_{ij}^{{u 0}}$, i.e.\ the DMI term is zero and both $\matp$ and $\mG$ only have components along the spin quantization axis, say the $z$ direction. Then, the $x$ and $y$ components of $\vec{\matp}$  and $\vec{\mG}$ vanish. Thus, $\vec{\matp}_i= (\mat{0}, \mat{0}, \matp_{i}^{z})$, $\mG^{x}=\mG^{y}=0$, $\mG^{0} = \frac{1}{2} (\mG^{\uparrow}+\mG^{\downarrow})$, $\mG^{z} = \frac{1}{2} (\mG^{\uparrow}-\mG^{\downarrow})$. The isotropic exchange parameter then reduces to
\begin{equation}
  \label{eq:LKAGA}
  J_{ij}^{iso} = \Im( A_{ij}^{00}- A_{ij}^{zz}).
\end{equation}
Defining $\Delta_i=\matp_i^{\uparrow}-\matp_i^{\downarrow} =2 | \vec{\matp}_i| = 2 \matp_i^z$ we obtain from Eq.\ (\ref{eq:LKAGA}) the LKAG \cite{Liechtenstein1987}
expression for isotropic exchange
\begin{equation}
  \label{eq:LKAG}
 J_{ij}^{iso} =  -\frac{1}{4\pi} \int_{-\infty}^{E_F}  \Im \Tr{\Delta_i\mG_{ij}^{\uparrow} {\Delta}_j\mG_{ji}^{\downarrow}}\ d\epsilon.
\end{equation}

\subsection{xyz average strategy}

We note that when all spins are oriented along one direction (e.g. $z$), the components with $u=z$ or $v=z$ in the $J$ tensor are non-zero if the variation with respect to the rotation is only kept to first order. The $xz$, $yz$, $zx$, $zy$, and $zz$ components of the anisotropic exchange, and the $D^z$ in the DMI cannot be obtained from a single calculation with magnetization along the $z$ direction. It should be noted that this does not affect the properties too much if the system stays close to the reference spin state, as these terms are small and higher order in the spin rotation angles. For example, in a AFM with small spin canting system, the parameters obtained from the AFM reference state can be used directly to get the canting angles. To determine the missing parameters, we rotate the whole system and lattice, (equivalently one can rotate the quantization axis) from $z$ to the $x$ and $y$ directions and apply the back-rotation to the DMI vectors. A weighted average over all three directions is taken: the inaccessible components are given a weight of 0 and the rest have the same weight. This small trick allows not only to obtain the full set of $\mat{J}^{ani}$ $\vec{D}$ vector components , but also to reduce the numerical noise. The DMI is usually much smaller than the isotropic exchange, such that its calculation is more numerically delicate. This issue arises in particular when Wannier functions are used whose symmetry is not guaranteed, and where the disentanglement procedure can introduce numerical errors. 

A direct approach to calculate the $D^{z}$ term is proposed in Ref.\ \cite{Mazurenko2005}, taking the spin-rotation perturbation to higher order. This is not implemented in TB2J, and would also be more sensitive to numerical noise. 

\subsection{Higher order terms}
It has been shown by several authors that the parameters from the Green's function method cannot be mapped exactly onto a bilinear $J$ tensor \cite{Lounis2010, Udvardi2003, Szilva2013} and higher order terms need to be considered, which include multi-spin interactions and the higher-order two-spin terms. In the simplest case, following Ref.~\cite{Szilva2013}, the Hamiltonian can be written in a biquadratic form as
\begin{equation}
  \label{eq:biquadratic}
  H^{Q} = -\sum_{i\neq j} J_{ij}^{\prime} \vec{S}_i \vec{S}_j  -\sum_{i\neq j} B_{ij} \left(\vec{S}_i \vec{S}_j\right)^2,
  \end{equation}
where $J_{ij}^{\prime}$ and $B_{ij}$ are determined by the $\mat{A}$ matrices as 
\begin{eqnarray}
J_{ij}^{\prime} &=& A_{ij}^{00} - 3A_{ij}^{zz},\label{eq:Jprime}\\
B_{ij}&=&A_{ij}^{zz}\label{eq:B}.
\end{eqnarray}

When $\vec{S}_i$ and $\vec{S_j}$ are close to their reference values, we have
\begin{eqnarray}
    J_{ij}&=&-\frac{d^2 H^Q}{d\vec{S}_i d\vec{S}_j} 
    = J^{\prime}_{ij} + 2 B_{ij} \vec{S}_i\vec{S}_j\\
    &\simeq& J^{\prime}_{ij} + 2 B_{ij} \vec{S}_i^{\text{ref}}\vec{S}_j^{\text{ref}}
    = J^{\prime}_{ij} + 2 B_{ij} = A_{ij}^{00} - A_{ij}^{zz}, \nonumber
\end{eqnarray}
which is equivalent to the $J_{ij}$ expression with only bilinear terms considered, and it shows that the effective bilinear $J$ term depends on the orientation of $S_i$ and $S_j$. 
This method was proposed in order to improve the description of the system when the deviation from the reference state is large~\cite{Szilva2013}. From our limited experience, however, the parameters produced by Eqs.\ (\ref{eq:Jprime}) and (\ref{eq:B}) are not always physical and their fit is more complex. 

\section{Implementation}
\begin{figure*}[htbp]
  \centering
  \includegraphics[width=0.95\textwidth]{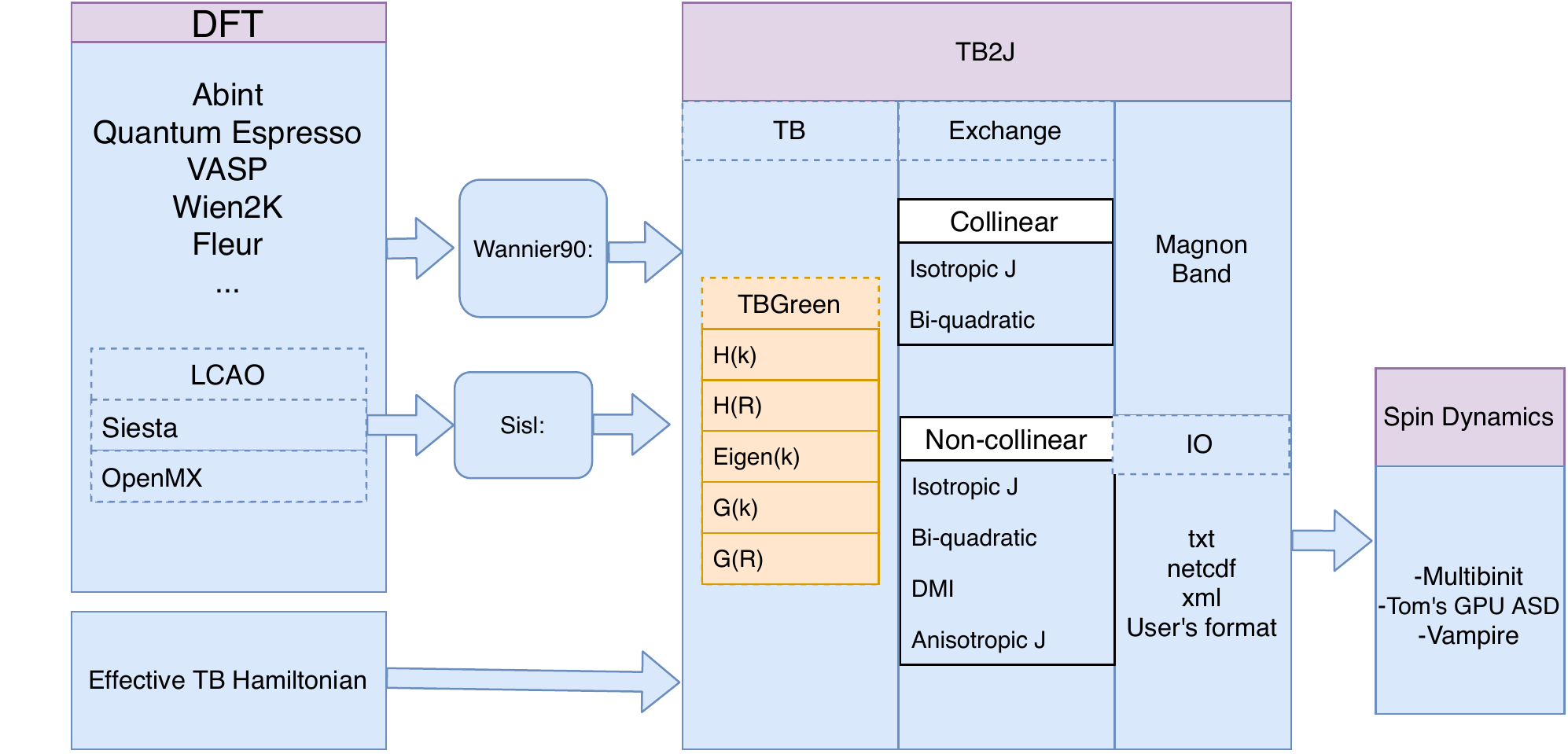}
  \caption{Schematic diagram view of TB2J workflow. }
  \label{fig:TB2Jdiag}
\end{figure*}

The TB2J package is implemented with three main modules, the tight-binding (TB) module, the exchange module, and the output module (see Fig.~\ref{fig:TB2Jdiag} for a schematic view). The TB module provides the interface with external DFT and Wannier function codes and calculates the Green's function. The exchange module uses the Green's function to calculate the magnetic interaction parameters. The output module writes these parameters to output files and provides a python API from which the magnon band structure can be calculated.

The TB module defines the classes of the TB model and the Green's function. The internal TB module can read the Wannier90 output data files to build the Hamiltonian. The WF's are assigned to the nearest atom with the crystal periodicity taken into account. A warning is issued if a WF is far away from any atom. The code utilizes the ``duck type'' feature: a class can be plugged in if the specific required methods are defined. The methods include the calculation of $H(\vec{k})$ and the eigenvalues and eigenvectors. Hence, external libraries can be wrapped easily. An interface to the sisl library \cite{zerothi_sisl} has been implemented, which includes the tight-binding model from Siesta and OpenMX. We also wrap directly the python based DFT code GPAW in LCAO mode instead of reading the Hamiltonian from the output files. 

Starting from the eigenvalues and eigenvectors, the Green's functions can be calculated without inverting the $\epsilon \mat{S}(\vk) - \mat{H}(\vk) $ matrix for every $\epsilon$. Instead, the Green's functions are calculated with $G(\vk, \epsilon) =\Psi(\vk) \left[ {\epsilon \mat{I} - Diag(E(\vk))} \right]^{-1}\Psi(\vk)^\dagger$, where $E(\vk)$ and $\Psi(\vk)$ are the eigenvalues and the eigenvector matrices, respectively. The real-space Green's function can then be calculated by Fourier transforming $G(k)$. 

Once the Green's functions are calculated, the exchange module calculates the magnetic interaction  parameters using Eqs.\ (\ref{eq:Jiso})-(\ref{eq:DMI}). The Hamiltonian and the Green's functions are first decomposed into their $\{0, x, y, z\}$ components. Then, the elements of the $4\times 4$ matrix, $\Tr{\mat{p}_i\mat{G}_{ij}^u \mat{p}_j\mat{G}_{ji}^v}$ for each $\epsilon$ are calculated and later integrated to obtain the $\mat{A}_{ij}$ matrix.

A contour integration method is used for the $\int^{E_F}d\epsilon$ integration. The range of the integration is $(E_{min}, E_F)$, where $E_{min}$ is either below the lowest band energy or chosen such that the orbitals below $E_{min}$ have only negligible interaction with those near $E_F$. By default, a semicircle path is used, which is centered at $(E_{min}+E_F)/2$ on the real axis and has a radius of $(E_F-E_{min})/2$, going through the upper half of the complex plane with $\Im(\epsilon)>0$. 

The magnetic interaction parameters are calculated from the  $\mat{A_{ij}}$ matrix. The full list of quantities that are available from TB2J is given in Table \ref{tab:quantites}.

\begin{table}[htbp]
  \centering
  {\setlength{\extrarowheight}{5pt}
  \begin{tabular}{cc}
    \toprule
   Quantity & Expression \\
    \midrule
            &    collinear case \\
    $J_{ij}$       & $-\frac{1}{4\pi} \int_{-\infty}^{E_F} d\epsilon \Im \Tr{\Delta_iG_{ij}^{\uparrow} \Delta_jG_{ji}^{\downarrow}}$      \\
    $J^\prime_{ij}$ & $Im( A_{ij}^{00}-3A_{ij}^{zz})$   \\
    $B_{ij}$ &  $A_{ij}^{zz}$ \\
        \hline
            &    non-collinear case \\
        $J_{ij}$  & $\Im(A_{ij}^{00}-A_{ij}^{xx}-A_{ij}^{yy}-A_{ij}^{zz})$   \\
    $\vec{D^{u}_{ij}}$ & $\Re (A_{ij}^{0u} - A_{ij}^{{u0}})$  \\
    $\mat{J}^{ani}$ & $ J^{ani,uv}= J^{ani,vu}= \Im (A^{uv}+A^{vu})$\\
    \hline
   \bottomrule 
  \end{tabular}
  }
  \caption{Summary of magnetic interaction quantities that are calculated within TB2J. All come from Ref.~\cite{Szilva2013} except the first (Ref. \cite{Liechtenstein1987}) and the DMI, which was first written in Ref. \cite{Antropov1997} }
  \label{tab:quantites}
\end{table}

\section{TB2J user instructions}
In this section, we describe the practical usage of TB2J. The users can also refer to the online documentation at \url{https://tb2j.readthedocs.io/en/latest/}, which is more detailed and regularly updated. We show how to install and use the code to calculate the magnetic interaction parameters using  body centered cubic (BCC) Fe, and hexagonal close packed (HCP) Co, BiFeO$_3$, and La$_2$CuO$_4$ as examples. 

\subsection{Installation}
 TB2J can be installed with a simple command if python and pip environments are pre-installed
\begin{lstlisting}[language=bash]
pip install TB2J
\end{lstlisting}
Alternatively, one can download the package from  \url{https://github.com/mailhexu/TB2J} and run
\begin{lstlisting}[language=bash]
python setup.py install
\end{lstlisting}
to install the package.

By default, TB2J only installs the hard (non optional) dependencies automatically. Hence, the sisl package, which is used to read the Hamiltonian from the Siesta or OpenMX output, needs to be installed separately using pip. Also, the GPAW-TB2J interface is through python directly, which requires the GPAW python package. 

\subsection{Computing the magnetic interaction parameters}
In this section we describe the TB2J procedures to compute the magnetic interactions, from an electronic structure calculation to the final output file. 
 
\subsubsection{Preparation of electronic structure and  tight-binding Hamiltonian}

To obtain the magnetic interaction parameters, the first step is to do a converged DFT calculation of a magnetic crystal, at the collinear or non-collinear level. Preferably, this calculation treats the magnetic ground state of the system, however, the magnetic interaction parameters can also be calculated for a different reference state. It should be noted that the spin quantization axis is presumed to be along the $z$ axis throughout TB2J.

The next step is to construct the tight-binding Hamiltonian. For DFT codes using non-local basis sets, like plane waves, the WFs can be constructed for any code which has an interface with Wannier90. All the spin-polarized orbitals of all atoms contributing to the magnetic interaction should be carefully selected to compute accurately the magnetic interactions interaction parameters. For example, in the case of a transition metal oxide the $d$ orbitals of the transition metal cation should be included, but also the oxygen $2p$ orbitals that are involved in the superexchange or DMI mechanism through hybridization with $d$ orbitals. This forms the minimal basis of orbitals to be included in the construction of the WFs but other orbitals might be important as well. The number of orbitals to be included is system dependent and should be checked by the user (convergence of the calculated magnetic interactions with respect to the number of Wannier orbitals). For example, in the case of SrMnO$_3$ the Mn-$3d$, O-$2p$ orbitals are necessary. The options to build the Wannier functions should be enabled in the DFT codes. For example, in Abinit, the ``prtwant 2'' and ``w90iniprj 2'' options (ABINIT version 9.x) are needed for the calculation of maximally localized Wannier functions (MLWFs). The input file for Wannier90 also needs to be present in the execution directory. THe quality of the Wannier function Hamiltonian is need to be checked. The magnetic interaction parameters are often meV or $\mu$eV, which requires the noise in the Hamiltonian to be lower. 

The rigid spin rotation on one site is performed by rotating all the spins of the WFs associated to a given atom, which requires the WFs to be centered on an atom or at least very close to it. TB2J uses the Wannier centers to decide which atom each WF ``belongs'' to. As a result, using the MLWFs  \cite{Marzari1997} might not always be the best choice. Projected WFs or selectively localized WFs~\cite{Wang2014}, which add a constraint on the Wannier centers, can be used instead. The Wannier centers for one atom might be located closer to a periodic copy of that atom than the original atom. To solve this problem, TB2J shifts the Wannier centers by using the transnational symmetry and modifies the Hamiltonian accordingly. The WF Hamiltonian and the center positions need to be output by Wannier90 using the ``write\_hr'' and the ``write\_xyz'' input flags.

For DFT codes based on LCAO basis sets, such as Siesta, the Hamiltonian is already localized and needs no further transformation. For calculating the parameters of the Heisenberg Hamiltonian only the localized DFT Hamiltonian and the overlap matrix need to be saved. For example, one can use the options ``CDF.Save=True'', ``SaveHS=True'', and ``Write.DMHS.Netcdf=True'' in Siesta (version 4.x) to enable the saving of these matrices.

\subsubsection{Running TB2J}
TB2J has two Python executables: wann2J.py and siesta2J.py, for calculating $J$ from Wannier90 and Siesta output, respectively. A similar script named openmx2J.py is in the TB2J\_OpenMX package. These scripts are designed to have a minimal user input where only the paths to the files containing the electron Hamiltonian information and defining the magnetic atom species need to be provided. 

\paragraph{With Wannier90}
The executable script wann2J.py can be used with the Wannier90 output files. For a non-collinear calculation, the spin up and spin down channel Wannier functions need to be present. In addition, the following parameters need to be specified:
\begin{itemize}
\item Whether the calculation is collinear or non-collinear, given by the --spinor option. TB2J assumes a collinear calculation by default, and the usage of --spinor specify that the calculation is non-collinear, where the Hamiltonian is in a spinor form.
\item the prefix to the paths of up and down Wannier functions, (--prefix\_up abinito\_w90\_up --prefix\_down abinito\_w90\_down, for non-collinear calculations, and --prefix\_spinor for
non-collinear calculations). The filename of the Hamiltonian is the prefix plus ``\_hr.dat". 
\item The posfile specifies the name
of a file containing the atomic structure and cell parameters. ASE file formats are also readable (the full list of which can be found on \url{https://wiki.fysik.dtu.dk/ase/ase/io/io.html}). It should be noted that some ASE formats cannot be used, e.g.\ xyz, because they do not contain the cell parameters which are required by TB2J. We also recommend using formats with the cell matrix rather than only the $(a, b, c, \alpha, \beta, \gamma)$, which will cause trouble if there are anisotropic or DMI terms as they are not rotationally invariant. 
\item The Fermi energy in units of eV.
\item The type of magnetic elements (symbol from the periodic table).
\end{itemize}
Here is an example for calculating the $J$s in the non-collinear case: 
\begin{lstlisting}[language=bash]
   wann2J.py --posfile abinit.in --efermi 5.8 --elements Fe --prefix_up abinito_w90_up --prefix_down abinito_w90_down 
\end{lstlisting}
In the case of a non-collinear, the WF Hamiltonian is in a single file and we need to specify the calculation type:
\begin{lstlisting}[language=bash]
   wann2J.py --posfile abinit.in --efermi 5.8 --elements Fe --prefix_spinor abinito_w90 --spinor 
\end{lstlisting}

\paragraph{With Siesta}
Only a minimal set of parameters is needed for Siesta: the filename of the input for the Siesta calculation. The other information needed, including whether the calculation has SOC enabled and the Fermi energy, are found in the Siesta results:
\begin{lstlisting}[language=bash]
 siesta2J.py --element Fe --input-fname='siesta.fdf' 
\end{lstlisting} 

\paragraph{With OpenMX}
The interface to OpenMX is distributed as a plugin to TB2J called TB2J\_OpenMX under the GPL license, which need to be installed separately, because code from OpenMX which is under the GPL license is used in the parser of OpenMX files. 
\begin{lstlisting}[language=bash]
pip install TB2J_OpenMX
\end{lstlisting} 
In the DFT calculation, the "HS.fileout on" options should be enabled, so that the Hamiltonian and the overlap matrices are written to a ''.scfout`` file. 
The necessary input are the path of the calculation, the prefix of the OpenMX files, and the magnetic elements:
\begin{lstlisting}[language=bash]
openmx2J.py --path ./ --prefix openmx --elements Fe
\end{lstlisting}

\paragraph{General options}
There are several tunable parameters, which the user usually does not need to specify. Among them, there are:
\begin{itemize}
\item nz: The number of steps in the path of the contour integration.
\item emin, emax: the integration lower and upper bounds, relative to the Fermi energy. The value of emin is automatically determined if not given, emax should be about 0 for  metallic systems, whereas it lies in the band gap for insulating systems.
\item rcut: the cutoff or max distance between spin pairs.
\end{itemize}
The full list of options can be accessed using \mbox{``--help''}.

As discussed in the previous section, the $z$ component of the DMI, and the $xz$, $yz$, $zx$, $zy$, $zz$ components of the anisotropic exchanges are non-physical, and an $xyz$ average is needed to get the full set of magnetic interaction parameters. In this case, scripts to rotate the structure and merge the results are provided, they are named TB2J\_rotate.py and TB2J\_merge.py. The TB2J\_rotate.py reads the structure file and generates three files containing the $z\rightarrow x$, $z\rightarrow y$ and the non-rotated structures. The output files are named atoms\_x, atoms\_y, atoms\_z. A large number of output file formats is supported thanks to the ASE library \cite{Larsen2017} and the format of the output structure files is provided using the \mbox{``--format''} parameter. An example for using the rotate file is:
\begin{lstlisting}[language=bash]
  TB2J_rotate.py BiFeO3.cif --format cif
\end{lstlisting}

The user has to perform DFT single point energy calculations for these three structures in different directories, keeping the spins along the $z$ direction, and run TB2J on each of them. After producing the TB2J results for the three rotated structures, we can merge the DMI results with the following command by providing the paths to the TB2J results of the three cases, e.g.: 
\begin{lstlisting}[language=bash]
  TB2J_merge.py BiFeO3_x BiFeO3_y BiFeO3_z --type structure
\end{lstlisting}
A new TB2J\_results directory is then made which contains the merged final results. 

\subsubsection{Output files}
\begin{lstlisting}[float=*ht, caption={An example of the output sections for BiFeO$_3$ with SOC enabled, calculated with spin along the $z$ axis. \label{lst:atomssection}},captionpos=b]

==========================================================================================
Information:
Exchange parameters generated by TB2J 0.2.8.
==========================================================================================
Cell (Angstrom):
 0.030   3.950   3.950
 3.950   0.030   3.950
 3.950   3.950   0.030

==========================================================================================
Atoms:
(Note: charge and magmoms only count the wannier functions.)
Atom_number      x         y         z     w_charge    M(x)      M(y)      M(z)
Bi1             0.2413    0.2413    0.2413    2.1878   -0.0010   -0.0005   -0.0045
Bi2             4.2060    4.2060    4.2060    2.1878    0.0005    0.0010    0.0045
Fe1             2.0165    2.0165    2.0165    6.1722   -0.0027   -0.0021    4.1151
Fe2             5.9812    5.9812    5.9812    6.1722    0.0021    0.0027   -4.1151
O1              5.5238    2.1558    3.9388    4.8807   -0.0005    0.0011   -0.0568
O2              6.0903    5.5389    3.9539    4.8807   -0.0011    0.0005    0.0568
O3              3.9388    5.5238    2.1558    4.8806   -0.0016   -0.0027   -0.0559
O4              3.9539    6.0903    5.5389    4.8806    0.0001   -0.0031    0.0562
O5              2.1558    3.9388    5.5238    4.8806    0.0031   -0.0001   -0.0562
O6              5.5389    3.9539    6.0903    4.8806    0.0027    0.0016    0.0559
Total                                        46.0038    0.0015   -0.0015   -0.0000

==========================================================================================
Exchange:
    i      j          R        J_iso(meV)          vector          distance(A)
----------------------------------------------------------------------------------------
   Fe2   Fe1   (  0,   1,   1) -26.7976   ( 3.934,  0.015,  0.015)  3.934
J_iso: -26.7976
[Experimental!] Jprime: -34.444,  B: -3.810
[Experimental!] DMI: ( 0.1590 -0.0996  0.0358)
[Experimental!]J_ani:
[[-0.026  0.002 -0.01 ]
 [ 0.002 -0.027 -0.05 ]
 [-0.01  -0.05  -7.62 ]]

\end{lstlisting} 

In the following we describe the output files which TB2J produces. By running wann2J.py or siesta2J.py, a directory with the name TB2J\_results will be generated, which contains the following output files:
\begin{itemize}
\item exchange.out: A human readable output file, which summarizes the results.
\item Multibinit: A directory containing output which can be read directly by the Multibinit code \cite{Gonze2020}.
\end{itemize}
The \textit{exchange.out} file contains three sections: cell, atoms and exchange. The cell section contains the lattice parameter matrix. The atoms section contains the positions, charges (for verification) and magnetic moments of the atoms: see Listing 1. 
 
Here, the charge and magnetic moment of each atom are only integrated with the WFs attached to this atom. As such they can differ from the quantities coming from the direct DFT output, as not all bands are used in the construction of WFs. The WF charges should be integers, for LCAO the values depend on the band energy cutoffs. In addition, the exclusion of very deep lying levels from the calculation of $J$ can also lead to deviations in the charges which might appear both for WFs and for LCAO. Another source of difference between the TB2J charges and magnetic moments and the DFT ones is the integration volume around the atoms, which is not necessarily the same. However, for localized $d$ and $f$ orbitals the magnetic moments should be close to their DFT counterparts, for TB2J to yield correct results for the parameters. Large differences between the TB2J and DFT values indicate that something may have gone wrong: either in in the contour integration $\int^{E_F} d\epsilon$ used in TB2J, or in the construction of the Wannier functions (incorrect wannierization process or too small WF basis set). Often, it comes from excluding an orbital that is important for the magnetic interaction in the studied system. In the case of metallic system, the Fermi energy might have to be slightly shifted with respect to the DFT reference due to different numerical method used in the integration of charge density.

Each pair of atoms is labeled by three parameters, the index $i$, $j$ and $R$, where $i$ and $j$ are the indices in the unit cell. The vector $\vec{R}$ specifies the cell the atom $j$ is translated to, i.e.\ the reduced positions of the two atoms are $\vec{r}_i$ and $\vec{r}_j+\vec{R}$, respectively. By default, the interaction is calculated within a supercell corresponding to the $k$-mesh. For example, with a $7\times 7 \times 7$ $k$-mesh, all $ij$ pairs will be produced for the spin labeled $i$ in the center cell of a $7\times 7 \times 7$ supercell. With the ``rcut'' flag, only the parameters for $ij$ pairs within a distance of rcut are calculated. The exchange parameters are reported as follows: magnetic atom $i$ connected with magnetic atom $j$, $R$ is the lattice vector between the unit cells containing $i$ and $j$, the value of $J$ for this pair of magnetic atoms in meV, the vector connecting them, and the distance between the pair of atoms. If SOC is enabled, the DMI and anisotropic $J^{ani}$ parameters are given in addition. The DMI vectors $\vec{D}$ and the anisotropic $J^{ani}$ are printed as vectors and matrices, respectively.  

Apart from the main \textit{exchange.out} file, TB2J delivers several other outputs, which provide the input for spin dynamics (SD) and Monte Carlo (MC) simulations. TB2J is interfaced with several SD and MC codes. It has native support to the Multibinit code delivered as part of the Abinit code since version 9.0 \cite{Gonze2020}. The  TB2J\_results/Multibinit directory contains the templates of input files for this code. One can usually run spin-dynamics with slight or no modification of these files. ``Experimental'' inputs are also generated for Vampire \cite{Evans2014} and Thomas Ostler's GPU-ASD code \cite{Digennaro2018}. The Output module of TB2J provides a versatile API, described in the online documentation (see Discussion section), which makes it easy to generate data for any other code. 

\subsubsection{Magnon band structure}
Once the $J$ have been obtained in real space from TB2J, the magnon dispersion curves $E_\mathrm{magnon}(\vec{q})$ can be obtained from the Fourier transform of $\mat{J}$ to the $\vec{q}$ space: 
\begin{equation}
  \label{eq:magnonband}
  \mat{J}(\vec{q})= \sum_{\vec{R}} \mat{J}(\vec{R}) e^{i\vec{q}\cdot\vec{R}},
\end{equation}
In the simple FM case this reduces to diagonalizing $\frac{4}{M}(J(0)-J(\vec{q}))$, where $M$ is the one site magnetic moment. The magnon band structure for more complicated magnetic configurations is more complex. Also, multiple types of magnetic sites cannot be treated at the moment. Both features will be added in a future version of TB2J.

With TB2J, the magnon band structure is obtained using the following command:
\begin{lstlisting}
TB2J_magnon --qpath GNPGHP --show
\end{lstlisting}
where the qpath option specifies the q-point path. If it is not specified, an automatically generate qpoint-path will be used from the information of the lattice structure by using ASE\cite{Larsen2017}. The details of the q-point path can be found on \url{https://wiki.fysik.dtu.dk/ase/ase/dft/kpoints.html}.  

A file, named ``exchange\_magnon.pdf'', containing the magnon band structure will be generated. The lowest energy eigenvector of $\mat{J}(\vec{q})$ is printed, from which, together with the $\vec{q}$ vector, one determines the ground-state spin configuration.

\subsection{Examples}
In order to demonstrate the usage of TB2J, we use five different examples, body centered cubic (BCC) Fe, hexagonal closed packed (HCP) Co, SrMnO$_3$, BiFeO$_3$, and La$_2$CuO$_4$. The BCC Fe and HCP Co systems are among the most studied magnetic materials and are used as standard benchmarks. The other three materials were chosen to represent a wide range of properties: SrMnO$_3$ as a prototype structure for superexchange, BiFeO$_3$ as a multiferroic material \cite{Catalan2009}, and La$_2$CuO$_4$ as a layered perovskite with a canted spin structure.    

\subsubsection{BCC Fe}
As a first benchmark, we calculate the magnetic interaction parameters for the probably most studied magnetic structure, BCC Fe. The isotropic exchange is calculated with TB2J-Siesta and compared with results from the TB-LMTO method in Ref.\ \cite{Pajda2001}, and KKR method in Refs.~\cite{Digennaro2018} and~\cite{Szilva2013}. We do not show TB2J-Wannier90 results because the localization procedure creates Wannier functions centered on bonds instead of atoms. Then, the mapping to a Heisenberg model is not well defined and the resulting exchange parameters are unphysical. 

The DFT calculations were performed with the Siesta code \cite{Soler2002} Max-1.0.13 version, with a double-zeta-polarized numerical atomic orbital basis set. The GGA-PBE \cite{Perdew1996} functional and the norm-conserving pseudopotentials from the pseudo-dojo \cite{Van2018} ``standard'' dataset in the psml \cite{Garcia2018} format were used. A $9\times 9\times 9$ $k$-point grid was used to sample the Brillouin zone. 

\begin{figure}
    \centering
    \includegraphics[width=0.45\textwidth]{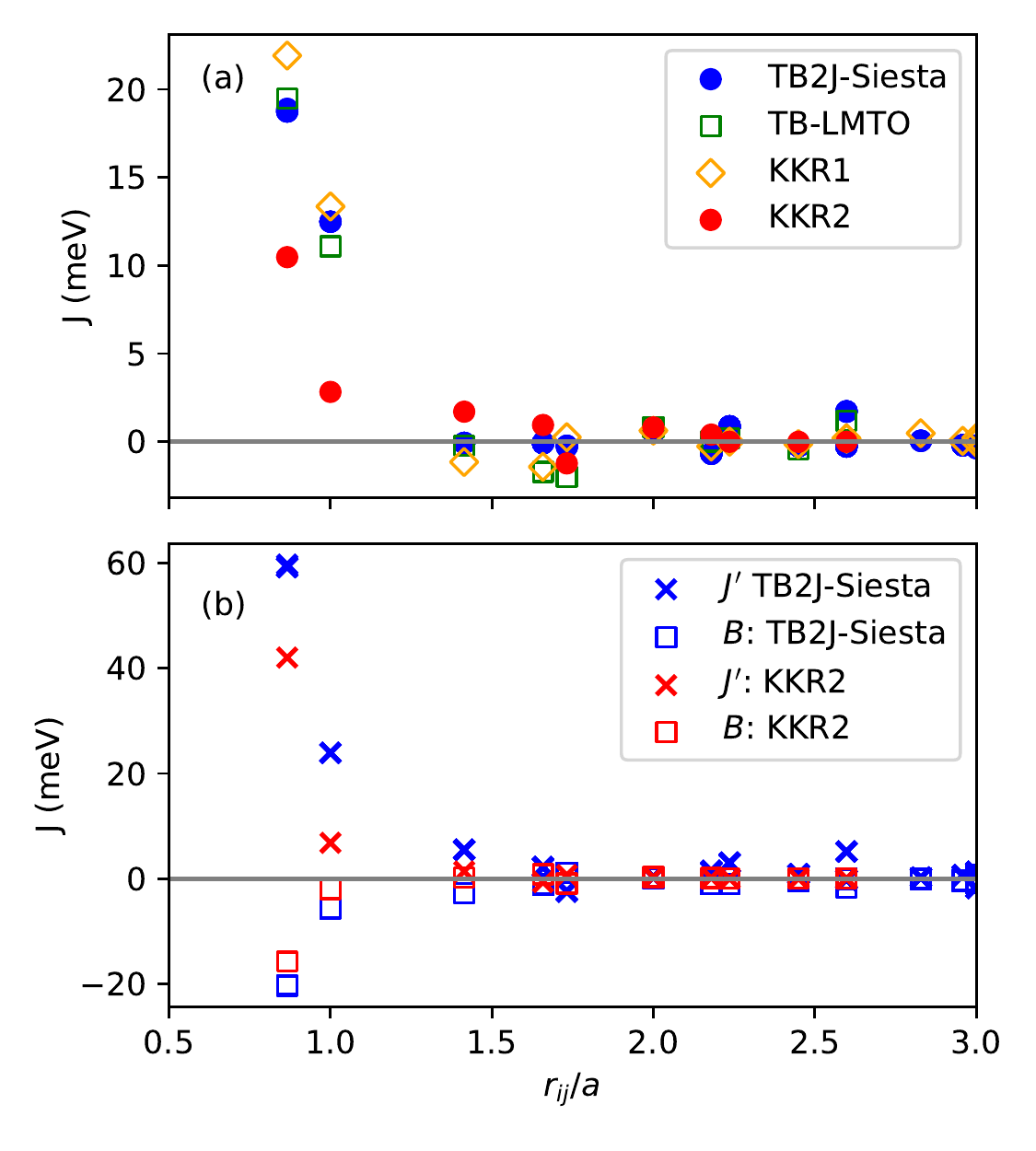}
    \caption{Exchange parameters for BCC Fe. (a) $J$ from TB2J-Siesta, TB-LMTO (Ref.\ \cite{Pajda2001}), and KKR (KKR1: Ref.\ \cite{Digennaro2018}, and KKR2: Ref.\ \cite{Szilva2013}). (b) The bilinear ($J^\prime$) and biquadratic ($B$) exchange of BCC Fe using TB2J-Siesta, and KKR-Green method (Ref.\ \cite{Szilva2013}). $r_{ij}$ is the distance between the Fe atom pairs, and $a$ is the cubic cell parameter.}

    \label{fig:bccFe}
\end{figure}

The results in Fig.~\ref{fig:bccFe} (a) show good agreement (within a few percent for the largest exchange parameters) between our calculation and those in Refs.\ \cite{Pajda2001} and~\cite{Digennaro2018} obtained with the GGA-PBE exchange correlation functional. The exchange parameters from the KKR method in Ref.~\cite{Szilva2013} are, however, significantly smaller. This might be attributed to a different density functional in the underlying DFT calculation, which is not reported in Ref.\ \cite{Szilva2013}. The decomposition of the exchange into bilinear and biquadratic terms is also calculated, see Fig.~\ref{fig:bccFe} (b), and compared with the KKR method from Ref.~\cite{Szilva2013}. The $J^\prime$ and $B$ from our calculation follow the same trend but are again larger than those in Ref.~\cite{Szilva2013}. 

\begin{figure}[htbp]
  \centering
  \includegraphics[width=0.47\textwidth]{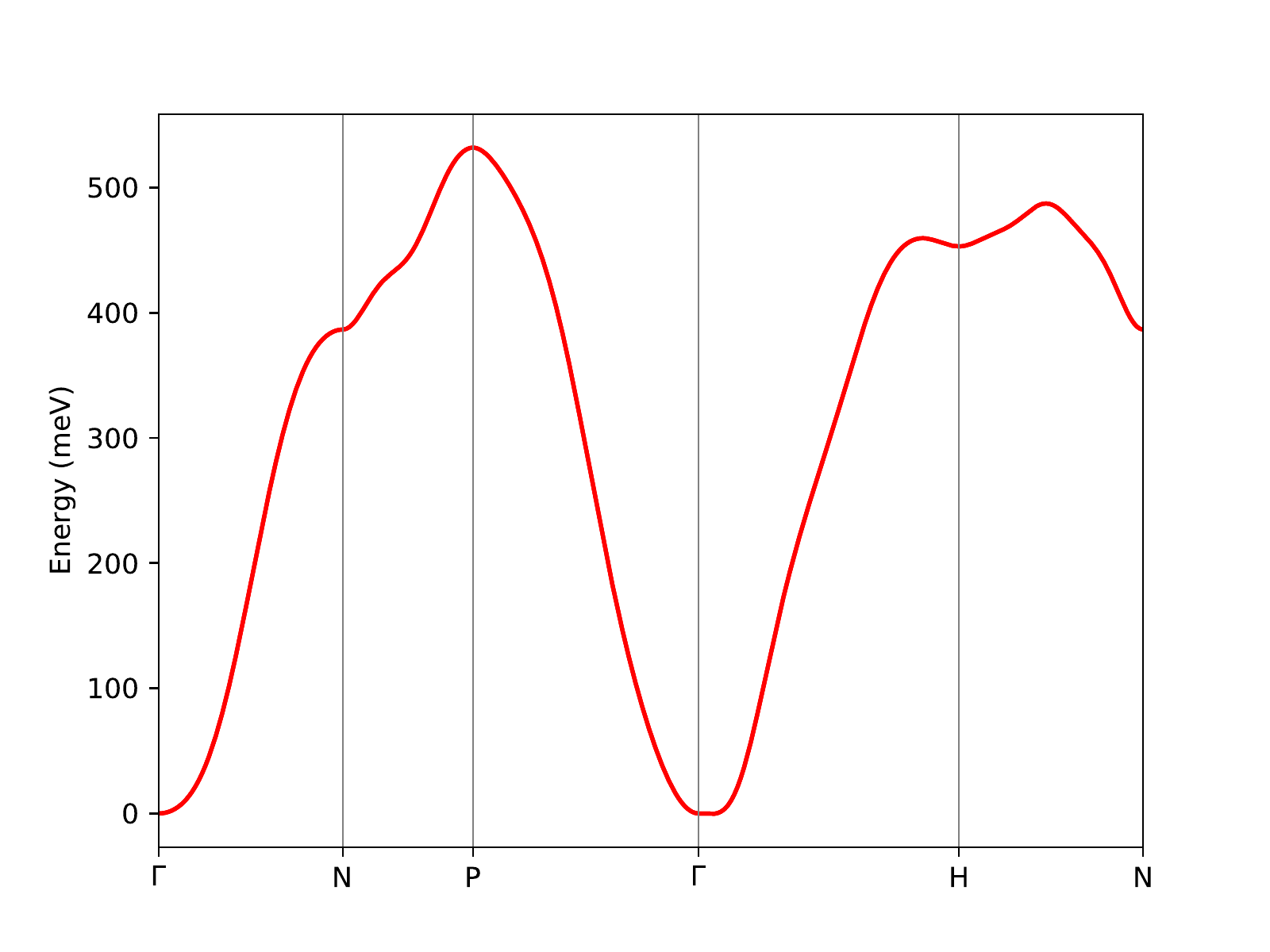}
  \caption{The magnon band structure of BCC Fe as obtained from TB2J. }
  \label{fig:bccFemagnon}
\end{figure}

Fig.~\ref{fig:bccFemagnon} shows an example of the magnon band structure of BCC Fe. As we can see, the ground state corresponds to a $q=\Gamma$ eigenvector, which identifies a ferromagnetic structure. The energy minimum and the corresponding eigenvector are also printed to the standard output:
\begin{lstlisting}
The energy minimum is at:
 q = [0. 0. 0.]
The ground state eigenvector is:
Fe1: [1.0 0.0 0.0]
\end{lstlisting}

\subsubsection{HCP Co}
\begin{figure}
    \centering
    \includegraphics[width=0.5\textwidth]{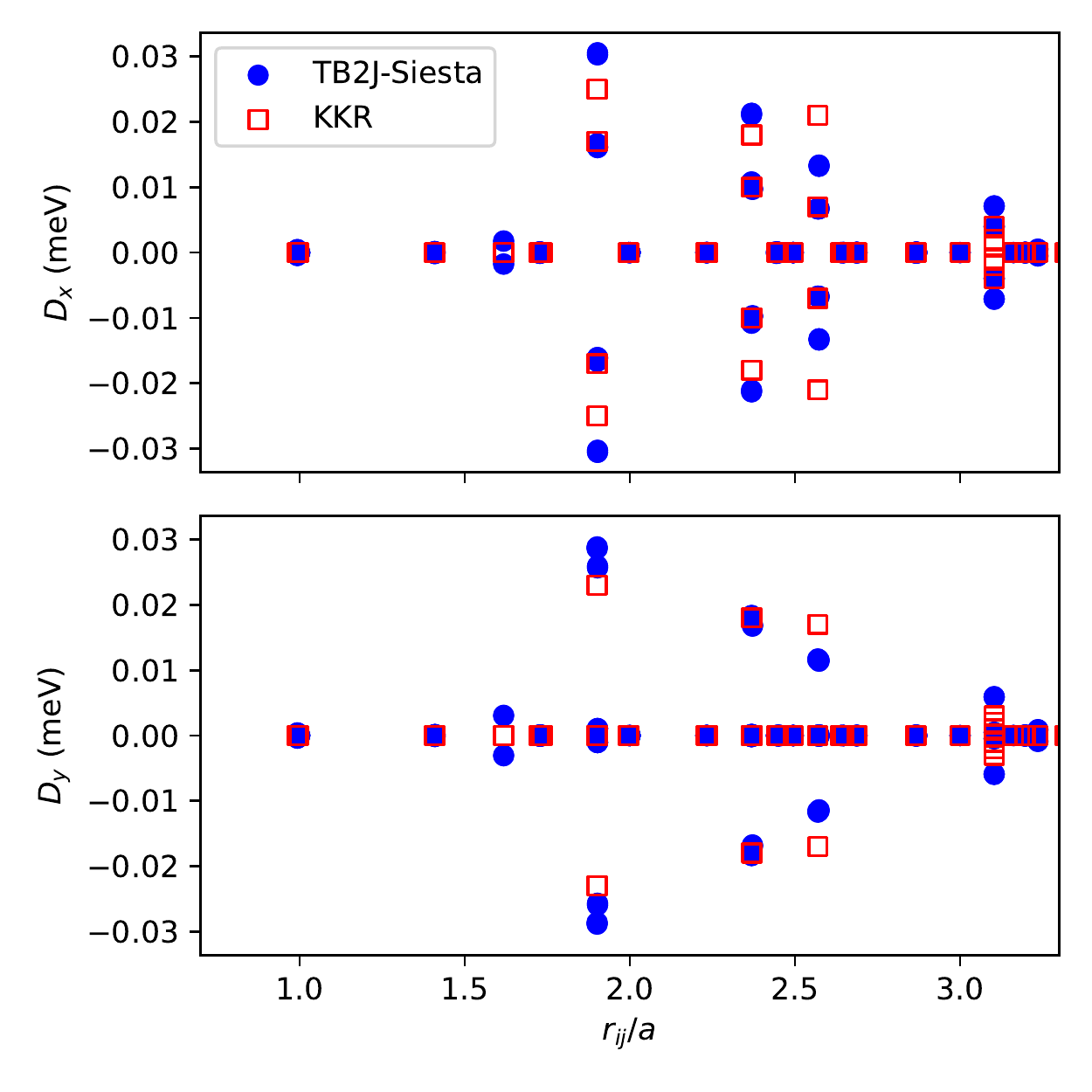}
    \caption{The DMI parameters of HCP Co calculated with TB2J compared with results calculated with the KKR-Green method reported in  Ref.\ \cite{Mankovsky2017}. }
    \label{fig:HCPCo}
\end{figure}

As a second example, we calculate the DMI in HCP Co with TB2J-Siesta and compare with the results from the KKR method of Ref.~\cite{Mankovsky2017}. A DZP LCAO basis set, together with a $9\times 9 \times 9$ $k$-point mesh, the GGA-PBE functional \cite{Perdew1996}, and the standard accuracy of v0.4 norm-conserving pseudopotentials from the Pseudo-Dojo  \cite{Van2018} are used. The SOC is enabled. The DMI in HCP Co is small (only of the order of 0.01~meV), which is a challenge to the accuracy of its calculation independent of the method used. The results in Fig.~\ref{fig:HCPCo} show good agreement between the two methods (within 10\% for most values). The $D_z$ component is not accessible when the $\vec{S}$ point along $z$.

\subsubsection{SrMnO$_3$}
In order to compare different DFT inputs, we calculate the exchange parameters for SrMnO$_3$ starting from either the Wannier90 or Siesta results. SrMnO$_3$ has a cubic perovskite structure with a lattice parameter of 3.81~$\AA$, and is a collinear antiferromagnet. The G-type antiferromagnetic ground state with a magnetic wave-vector of $(1/2, 1/2, 1/2)$ is taken as the reference configuration in our DFT calculations. It requires a 10-atom supercell expanded along the [111] direction.  

The DFT calculation for constructing the Wannier functions is performed with Abinit v9, using a plane-wave basis set with a cutoff of 30~Ha. The projected augmented wave (PAW) \cite{Torrent2008} potentials are from the ``JTH'' dataset \cite{Jollet2014} v1.2. The Mn $3d$ and O $2p$ orbitals are considered in the construction of the maximally localized Wannier functions from which the  tight-binding Hamiltonian is calculated. In the Siesta calculation, we use a DZP LCAO basis set for all atoms and the pseudodojo v0.4 norm-conserving pseudopotentials \cite{Van2018}. In both DFT calculations, we use the PBEsol \cite{Perdew2008} functional to approximate the exchange-correlation, with a Hubbard $U$ \cite{Anisimov1991} of 3~eV on the Mn atom. The SOC is not enabled. A $9\times 9 \times 9$ $k$-point mesh is used for the integration of the Brillouin zone. 

We show the calculated magnetic interaction parameters with both Abinit+Wannier90 and Siesta in Table \ref{tab:SMO} and compare them with total energy results from Ref.\ \cite{Lee2011b}. We can see that all three methods yield similar qualitative results with a large and negative parameter for first nearest neighbours and  small parameter for second and third neighbour interaction. 

\begin{table}
    \centering
    \begin{tabular}{ccccc}
    \toprule
    & $|S|$ &1NN & 2NN & 3NN \\
    \midrule
TB2J-W90-Abinit &2.81 &-8.16 &-0.36 &-0.02 \\
TB2J-Siesta & 2.85& -7.70 &-0.02 &0.11  \\
TE \cite{Lee2011b} &  & -6.98 & -0.36 & -0.01\\
\bottomrule
\end{tabular}
    \caption{The exchange interaction parameters of SrMnO$_3$, calculated with TB2J-Wannier90-Abinit (TB2J-W90-Abinit) and TB2J-Siesta, compared to the results of the total energy (TE) method in Ref.~\cite{Lee2011b}. $|S|$ and the exchange parameters are given in units of $\mu_B$ and meV, respectively.  Ref.~\cite{Lee2011b} uses a different convention and a $|S|^2 = (3/2)^2$ is multiplied to the values in it to convert the values to the convention of the present paper. They also use a slightly smaller $U$(Mn)=2.7~eV.}
    \label{tab:SMO}
\end{table}

\subsubsection{BiFeO$_3$}

BiFeO$_3$ has a cubic-perovskite-like structure, in which the Bi atoms are at the corners, the O atoms are at the face centers and the Fe ions are in the center of the cube. The $R3c$ state structure has two units of BiFe$O_3$, and can be seen as the cubic structure plus the ``$a^-a^-a^-$'' rotation \cite{Glazer1972} of the O octahedra plus the polar distortion along the [111] direction. The DMI is related to the local inversion symmetry breaking due to these distortions \cite{weingart2012}. 

The DFT results are from Siesta Max-1.0-13 version, with a DZP LCAO basis set for all atoms. We used the full-relativistic norm-conserving pseudopotentials from the Pseudo-Dojo "standard" dataset, with the GGA-PBE \cite{Perdew1996} density functional, corrected by a Hubbard term\cite{Anisimov1991} with $U(Fe)=4$,  and a $7 \times 7 \times 7$ $k$-point mesh. The SOC is enabled for the calculation of the anisotropic and DMI terms.


The calculated magnetic interaction parameters are given in Table \ref{tab:BFO}. The first nearest neighbor exchange and DMI are compared with both a DFT total energy method from Ref.\ \cite{Xu2019} and experimental result from Ref.\ \cite{Matsuda2012}. In Ref.\ \cite{Xu2019}, the DFT+U \cite{Anisimov1991} correction is used with $U(Fe)=4$. 
The results show that the isotropic exchange and DMI parameters from TB2J-Siesta and the total energy method are close. The first NN $J$ and DMI also agree well with the experimental value. 

\begin{table*}
\begin{center}
  \begin{tabular}{rcccccccr}
    \toprule
    Method & $n^{th} NN$ & $(R_{ij})$ & $(J^{iso})$ & $(J_{xy})$ & $(J_{yz})$ & $(J_{xz})$ & $(D_x,  D_y,  D_z )$ & $|D|$ \\
    \midrule
    TB2J-Siesta &1 & (1,0,0) & -26.79 & 0.003 & 0.001 & -0.008 & (0.158, -0.093, 0.325) & 0.373\\
    Exp\cite{Matsuda2012}& & &-20.25  &&&&&  0.507\\
    TE\cite{Xu2019}& & &-18.296 & & & & (0.131, -0.087, 0.363) & 0.396  \\
    \hline
    TB2J-Siesta&2 & (1,-1,0) & -0.742 & 0.004 & -0.001 & -0.008 & (-0.003, -0.009, -0.003) & 0.010\\
         TE\cite{Xu2019}& & &-0.603 & & & & (-0.003, -0.006, -0.006) & 0.009 \\
\hline    
      TB2J-Siesta&3 & (1,1,0) & -0.023 & -0.000 & 0.012 & -0.013 & (-0.036, 0.003, 0.034) & 0.049 \\
         TE\cite{Xu2019}& & &-0.009 & & & & (-0.000, -0.006,-0.012) & 0.013 \\
    \bottomrule
  \end{tabular}
  \label{tab:BFO}
  \caption{The magnetic interaction parameters (in meV) of BiFeO$_3$ compared with total energy (TE) method and experimental (Exp) measurement. In Refs.\ \cite{Xu2019} and \cite{Matsuda2012}, a different convention is used, a $-|S|^2/2=-(5/2)^2/2$ is multiplied to adapt to the convention used in this work.}
\end{center}
\end{table*}

\subsubsection{La$_2$CuO$_4$}
\begin{figure}
    \centering
    \includegraphics[width=0.4\textwidth]{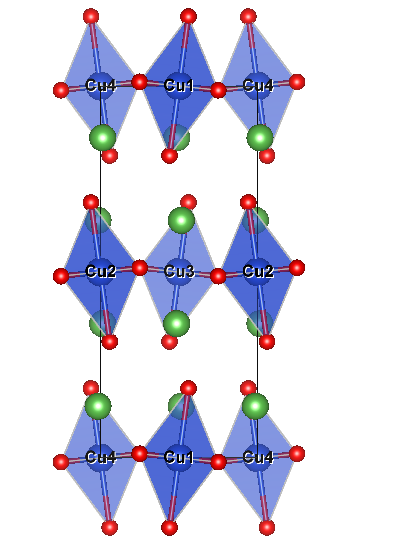}
    \caption{The structure of La$_2$CuO$_4$. La atoms are depicted in green, Cu atoms in blue, and O atoms in red. The figure is in the $x-z$ plane with $x$ and $z$ in the horizontal and vertical direction.}
    \label{fig:LaCuOstructure}
\end{figure}
In this section we show an example of calculating the magnetic interaction parameters for a spin-canting material, namely La$_2$CuO$_4$. It has a layered structure where the Cu-O octahedra are separated by the La-O planes, see Fig.\ \ref{fig:LaCuOstructure}. There is an out-of-plane displacement for each of the O ions in the Cu-O plane ($xy$ plane) due to the $a^-b^-$ rotation of the Cu-O octahedra.

The DFT calculations were performed with OpenMX \cite{Ozaki2005} using the numerical pseudo-atomic orbital (PAO)  basis set, with La s3p2d2f1, Cu s3p2d1, and O s3p2d1 PAOs as provided in the OpenMX database \cite{Ozaki2004}  v2019. A $6\times 6 \times 4$ $k$-point grid and the GGA-PBE \cite{Perdew1996} functional with a Hubbard $U$ correction \cite{Liechtenstein1995} ($U$(Cu)=8 eV, $J$(Cu)=0.8 eV) are used. The primitive cell of 4 $\times$ La$_2$CuO$_4$ with an antiferromagnetic (AFM) spin alignment in the $xy$ plane is taken as the reference DFT state. The isotropic exchange and DMI parameters are calculated. The interactions with the first nearest neighbors of one Cu atom are listed in Table~\ref{tab:LaCuO}.  The results are compared with the LMTO results from Ref.~\cite{Mazurenko2005}, in which a $U$(Cu)=10 eV and $J$(Cu)=1 eV are used. The exchange interaction shows an AFM alignment. The sum of the $\vec{D}$ has a non-zero component along $x$, which results in a net magnetic moment perpendicular to $x$. The angle of rotation can be estimated as $|\sum_{j\neq i}{\vec{D}_{ij}}|/{|2\sum_{j\neq i} J_{ij}|}$, which is about $0.8\times 10^3$, close to $0.7\times 10^3$ in Ref.~\cite{Mazurenko2005}. 


\begin{table*}
    \centering
\begin{tabular}{ccccc}
\toprule
& \multicolumn{2}{c}{TB2J-OpenMX} & \multicolumn{2}{c}{LMTO~\cite{Mazurenko2005}} \\
$\vec{R}_{0j}$ & J & $\vec{D}$ & J & $\vec{D}$ \\
\midrule
(-0.49,0.5,0) & -18.60 & (-0.023, 0.025,0.001 )  & -14.576 & (-0.020, 0.032, 0.005)\\
 (0.49, 0.5, 0) & -18.60 & (-0.023, -0.025, -0.001) & -14.576 & (-0.020, -0.032, -0.005)\\
(0.49, -0.5, 0) & -18.60 & (-0.023, 0.025, 0.001) & -14.576 & (-0.020, 0.032, 0.005)\\
(-0.49, -0.5, 0) & -18.60 & (-0.023, -0.025, -0.001) & -14.576 & (-0.020, -0.032, -0.005)\\
(0, 1, 0) & 5.82 & (0,0,0) & 2.071 & (-0.002, 0,0) \\
(0, -1, 0) & 5.82 & (0,0,0) & 2.071 & (-0.002, 0,0) \\
(-0.98, 0, 0) & 4.96 & (0,0,0) & 1.943 & (0.007, 0,0) \\
(0.98, 0, 0) & 4.96 & (0,0,0) & 1.943 & (0.007, 0,0) \\
Total & -52.84 & (-0.092,0,0) & -48.476 & (-0.070, 0, 0) \\
\bottomrule
\end{tabular}
    \caption{The isotropic exchange and DMI parameters for La$_2$CuO$_4$. The neighbors of the Cu1 atom shown in Fig.~\ref{fig:LaCuOstructure} is shown. The $\vec{R}_{0j}$ is the vector connecting the neighbors with Cu1. are in the unit of lattice vectors.  The units of $J$ and $\vec{D}$ are meV and are in laboratory coordinates. The convention used in Ref~\cite{Mazurenko2005} is different from the one in the present work by a minus sign.}
    \label{tab:LaCuO}
\end{table*}


\subsection{Spin dynamics}
\begin{figure}[htbp]
  \centering
  \includegraphics[width=0.43\textwidth]{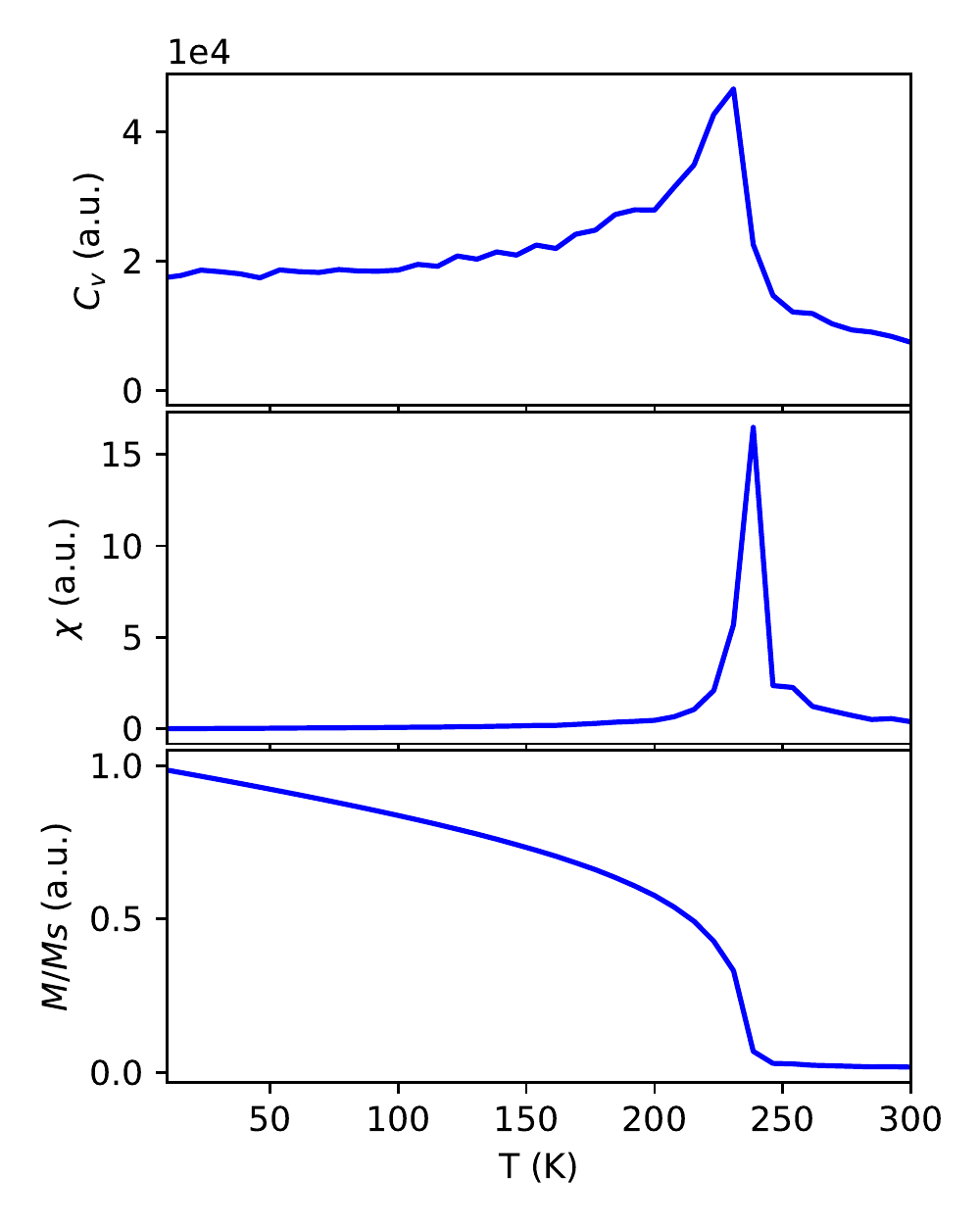}
 
  \caption{The volume specific heat, $C_V$ (top), the susceptibility, $\chi$ (middle), and the ratio of magnetic moments, $M/M_S$ (bottom), as functions of $T$ in SrMnO$_3$ as obtained with MULTIBINIT spin dynamics. The units are atomic units.}
  \label{fig:SMOSD}
\end{figure}
We use a simple example of calculating the N\'eel temperature ($T_N$) of SrMnO$_3$ to show the usage of the TB2J results in spin dynamics with the MULTIBINIT code. In the TB2J output, the templates of the Multibnit input files (\mbox{mb.files} and \mbox{mb.in}) are already generated. To calculate $T_N$, we simply adjust a few parameters, e.g.\ the size of the supercell and the temperature range and steps. Then, by running 
\begin{lstlisting}
multibinit < mb.files
\end{lstlisting}
the spin dynamics at various temperatures are performed. For the parameters, we employ the ones from the TB2J-Siesta calculation. A file named mb.out.varT contains the magnetic volume specific heat, $C_V$, the susceptibility, $\chi$, and the ratio of the magnetic moment to the saturated value, $M/M_S$. We plot the results as functions of temperature,  $T$, in Fig.~\ref{fig:SMOSD}, from which we can see that $T_N$ is about 240 K, which is close to the experimental value of about 233 K. \cite{Chmaissem2001}. 

\section{Discussion}

\subsection{Advantages of the Green's function method}
The Green's function method implemented in TB2J has several advantages over the total-energy mapping methods. It is more robust, and only requires a DFT calculation for a single spin configuration, typically the ground-state one. Therefore, it avoids the problem of converging DFT calculations for magnetic structures which are much higher in energy or even unstable. For example, a structure with an insulating ground state may become metallic in some spin configurations (often the FM phase is metallic while the AFM is not), or electrons become delocalized, which modifies or destroys the magnetic character. The convergence of non-collinear cases with SOC can be especially difficult in DFT, and avoiding several such calculations is a strong advantage. Spin dynamics at moderate temperatures typically stay close to the ground state, and fitting to extreme configurations, e.g.\ Stoner excitations, will produce worse dynamics. Even up to the Curie temperature, perturbative calculations typically perform well.

The Green's function method is most accurate around the magnetic phase for which the DFT calculation has been performed. This is in contrast with the total-energy methods in which the parameters are averaged over diverse magnetic configurations. However, experience shows that the two methods yield similar results for the parameters for many systems.

An important caveat is that the parameters from total-energy fitting methods are supercell-dependent. For example, if the supercell only contains the second nearest neighbours (2NN), the magnetic interaction of longer range is wrapped into the 1NN and 2NN parameters, and large cells have to be built to check the convergence of the fitting, which is not always numerically feasible. In the case of TB2J, the building of supercells is avoided. This makes the method suitable for large scale calculations, where the exchange magnetic interaction parameters can be calculated at any distance in real space, provided the $k$-point sampling is dense enough.

TB2J can be used either in conjunction with WFs or with LCAO codes. For WFs, TB2J works almost as a black box. Indeed, the whole procedure - from DFT data to the calculation of magnetic interaction parameters - requires little or no user input. In the case of LCAO codes, TB2J also runs with almost no extra input. As a result, TB2J can be easily integrated into high-throughput workflows for automatic high-throughput calculations of the magnetic interactions, using packages like ASE \cite{Larsen2017}, Abipy \cite{Gonze2020}  or Aiida \cite{Pizzi2016}, since TB2J can work as a python library. 

The construction of MLWFs was a sophisticated task in the past, since input parameters would vary for each structure and had to be tuned by hand. However, the situation has been largely changed thanks to the recent development of several methods, in particular the selected columns of density matrix (SCDM) algorithms \cite{Damle2015, Damle2018}. With these advances, fully automated building of WFs has been demonstrated \cite{Vitale2019}. 

The output of TB2J is directly usable by spin simulation codes. Therefore, the full procedure from the DFT exchange calculation to the spin system simulation can be easily automated.

\subsection{Limitations}
In the limit where the method implemented in TB2J applies (rigid spin rotation approximation, magnetic moments localized on the atoms, etc.) and that the limitations of the DFT for magnetic systems (selection of a correct enough exchange-correlation functional, choice of the values of DFT+U parameters, etc \cite{varignon2019, himmetoglu2014, bousquet2010}) the  calculation of the DMI and anisotropic exchange parameters through the Wannier basis set might still be problematic. 
The user should always verify that the electronic band structure obtained with the Wannier functions is in good agreement with the DFT one (choice of good disentanglement energy window, choice of projectors, etc). However, even if that is the case, the Wannierization process itself can introduce noise of the order of magnitude of a few $\mu$eV. For small quantities, like the DMI or anisotropic exchanges, which are of the same size as the noise, the parameters cannot be determined with sufficient resolution.
The problem occurs, in particular, when strong disentanglement of bands is necessary, i.e.\ when no gap is present between selected bands. It can be reduced in cases where the bands of interest are not entangled too much with higher unoccupied bands by including more unoccupied bands in the Wannerization process. Therefore, a validation of the quality of Wannier function Hamiltonian should be performed to make sure that the calculation is meaningful, by comparing the band structures from DFT and from Wannier function based Hamiltonian.
We have tested using the symmetry-adapted Wannier functions \cite{sakuma2013} which, however, does not fully solve the problem. It appears that the calculation of very small parameters using Wannier functions poses a challenge which hopefully future developments in the Wannierization methods can address by reducing the noise introduced in the disentanglement procedure.

At last, the Heisenberg model might not be valid for all circumstances, for instance, when spins are itinerant or when the non-bilinear magnetic interactions are non-negligible. The rotation of the spins is described in a basis set, either Wannier functions, or numerical atomic orbitals. In some cases, especially when the spins are not localized at or close to the centers of the basis functions, the result can be basis set dependent.



\subsection{Code availability}
The code is freely available under the BSD~2 clause license and can be found at  \url{https://github.com/mailhexu/TB2J/}. Documentation is provided at  \url{https://tb2j.readthedocs.io/en/latest/}. The interface to OpenMX is in a separate package, TB2J\_OpenMX (\url{https://github.com/mailhexu/TB2J-OpenMX}), under the GPLv3 license, which use TB2J as a library. Contribution to the code is welcome. In particular, we are happy to integrate interfaces to additional codes, be it on the input side (other first-principles or tight-binding codes), or the output side (e.g.\ outputs to atomic spin simulation packages). 

\section{Conclusions and perspectives}
In this paper we present the TB2J python package which calculates the magnetic interaction parameters from Wannier and LCAO Hamiltonians, using a Green's function method. The isotropic and anisotropic exchange, and the DMI can be systematically calculated by TB2J. The code can use results from a large number of first-principles DFT codes, either through the Wannier90 interface (for plane wave codes but also other basis sets) or directly from LCAO codes (SIESTA, GPAW, and OPENMX). TB2J requires a minimal number of input parameters, such that it can be easily integrated in high-throughput workflows. One of the most appealing features of TB2J is the requirement of only one unit cell DFT calculation (or eventually 3 for numerical averaging of $x$, $y$, $z$ spin orientations) to evaluate the magnetic interactions at any distance between the magnetic atoms. We hope this development will simplify the life of future users, in their calculations of the magnetic interactions from first-principles ingredients, and enable larger scale and more accurate micromagnetics calculations, exploring novel physical phenomena.

Future versions will include a wider range of interfaces with DFT and ASD codes, as well as expanded feature sets, including single ion anisotropy, magnon band structures for more complex systems.
A promising avenue is the calculation of higher order parameters (3 spin, 4 spin), which has been explored only very sparsely in the literature (e.g.\ Refs. \cite{fedorova2015,Ostler2017,Mankovsky2020,Hoffmann2020}).

\section*{Acknowledgements}
The authors thank Yajun Zhang, Alireza Sasani, Jorge Pilo González, and Zachary Romestan for the testing of the code, Thomas Ostler, Bertrand Dup\'e and Phivos Mavropoulos for explanations about the limits and intricacies of fitting the Heisenberg model. 

This work has been funded by the Communaut\'e Fran\c{c}aise de Belgique (ARC AIMED G.A. 15/19-09).XH thanks the support by the EU H2020-NMBP-TO-IND-2018 project "INTERSECT" (Grant No. 814487).
EB thanks the FRS-FNRS for support, as does MJV for an ``out'' sabbatical grant to ICN2 Barcelona in 2018-2019.
The authors acknowledge the CECI supercomputer facilities funded by the F.R.S-FNRS (Grant No. 2.5020.1) and the Tier-1 supercomputer of the F\'ed\'eration Wallonie-Bruxelles funded by the Walloon Region (Grant No. 1117545).
Computing time was also provided by PRACE-3IP DECI grants 2DSpin and Pylight on Beskow (G.A. 653838 of H2020).

\bibliographystyle{cpc}
\interlinepenalty=1000
\bibliography{mybib.bib}
\end{document}